\documentclass[prb,twocolumn,showpacs]{revtex4}
\usepackage{bm}
\usepackage{graphicx}
\usepackage{amsmath}
\usepackage{eufrak}
\usepackage{color}
\newcommand{\nix}[1]{}
\begin{document}

\title{Quantum ratchet effects induced by terahertz radiation in GaN-based
two-dimensional structures}

\author{W.~Weber,$^1$ L.\,E.~Golub,$^2$  S.\,N.~Danilov,$^1$ J.~Karch,$^1$
C.~Reitmaier,$^1$ B.~Wittmann,$^1$ V.\,V. Bel'kov,$^2$
E.\,L.~Ivchenko,$^2$ Z.\,D.~Kvon,$^3$ N.\,Q.~Vinh,$^4$
A.\,F.\,G.~van~der~Meer,$^4$ B.~Murdin,$^5$ and
S.\,D.~Ganichev$^{1}\footnote{e-mail:
sergey.ganichev@physik.uni-regensburg.de}$}
\affiliation{$^1$  Terahertz Center, University of Regensburg,
93040 Regensburg, Germany}
\affiliation{$^2$A.F.~Ioffe Physico-Technical Institute, Russian
Academy of Sciences, 194021 St.~Petersburg, Russia}
\affiliation{$^3$ Institute of Semiconductor Physics, Russian
Academy of Sciences, 630090 Novosibirsk, Russia}

\affiliation{$^4$ FOM Institute for Plasma Physics ``Rijnhuizen'',
P.O. Box 1207, NL-3430 BE Nieuwegein, The Netherlands}

\affiliation{$^5$University of Surrey, Guildford, GU2 7XH, UK}


\begin{abstract}
Photogalvanic effects are observed and investigated in wurtzite
(0001)-oriented GaN/AlGaN low-dimensional structures excited by
terahertz radiation. The structures are shown to represent linear
quantum ratchets. Experimental and theoretical analysis exhibits
that the observed photocurrents are related to the lack of an
inversion center in  the GaN-based heterojunctions.
\end{abstract}


\maketitle

\section{Introduction}

In recent years, physics of quantum ratchets draw a growing
attention. In mechanics, a ratchet is a device that is used to
restrict motion in one direction while permitting it in another.
In general, one means by a ratchet any sort of asymmetric
potential, or a potential lacking a center of the spatial
inversion. At the turn of 1980/90's it was understood that
unbiased nonequilibrium noncentrosymmetric systems can generate
transport of particles, and this conception has roots in different
fields of physics, chemistry and biology. In mechanical,
electronic, optical and biological systems, a particle, classical
or quantum, charged or neutral, propagating in a periodic
potential with broken centrosymmetry and subjected to an {\it ac}
force exhibits a net {\it dc} macroscopic flow. Examples are
electrons in solids, Abrikosov vortices in superconductors and
biological motor proteins.\cite{applphys} One of implementations
of the ratchet phenomenon is a Linear Photo-Galvanic Effect
(LPGE). It represents a generation of  a {\it dc} electric current
under absorption of linearly polarized light in unbiased crystals
of piezoelectric classes. Glass et al.\cite{glass} were the first
to attribute a photoinduced current observed in ferroelectric
LiNbO$_3$ to a novel photogalvanic effect and propose the first
correct model for its microscopic qualitative interpretation. A
consistent quantitative theory of LPGE was developed by Belinicher
and Sturman followed by other theorists, see
Refs.~\onlinecite{SturmanFridkin,10}.

Noncentrosymmetric bulk semiconductors and heterostructures are
natural quantum ratchets and the further studies of LPGE in these
systems allow one to elucidate the whole problem of the ratchet
effect. In particular, a current of the LPGE consists of two
contributions, ballistic and shift, which as a rule are comparable
in the order of magnitude.\cite{BelIvchStur}
It is important to note that similar two contributions to the
electric current, called the skew-scattering and side-jump
currents, do exist in the anomalous Hall effect, see the recent
review by Sinitsyn.\cite{sinitsyn2}
Another important point is that one should include into the light
absorption process contributing to the ballistic current an
additional simultaneous scattering from a static defect or a
phonon, as compared to the light-induced optical transition
leading to the shift current. In GaN-based structures first
observation of LPGE has been reported quite
recently.\cite{APL05,ICPS06,HeAPL2007,TangApl2007,ChoPRB2007,SSC07}
Here we present results of detailed experimental and theoretical
investigation of LPGE in heterostructures based on GaN and its
alloys with AlN. The commercial fabrication of blue and green LEDs
have led to well established technological procedures of epitaxial
preparation of these heterostructures and initiated a great
activity on investigations of their properties.\cite{Nakamura} The
photogalvanics serves as a solid bridge between transport and
optics and, therefore, reveals both transport and optical
properties of the systems under study.

\section{Samples and experimental methods}

\begin{figure}[t]
\includegraphics[width=0.7\linewidth]{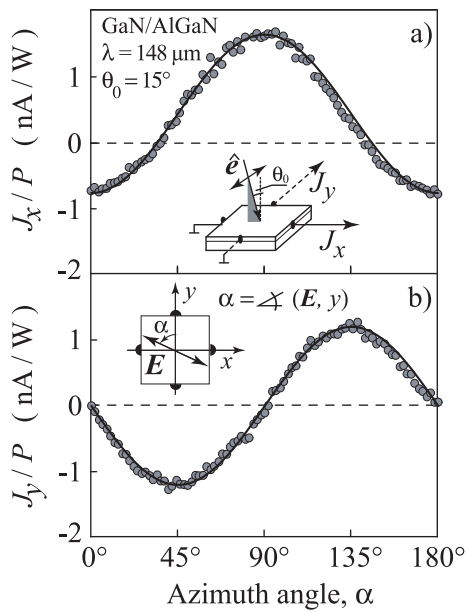}
\caption{Photocurrent as a function of angle $\alpha$ measured at
room temperature at oblique incidence ($\theta_0 = 15^\circ$) in
the a)~longitudinal ($J_x$) and b)~transverse ($J_y$)  geometries.
Photocurrent is excited by linearly polarized radiation with
wavelength $\lambda$~=~148~$\mu$m and power $P \approx 5$~kW. Full
lines are fits to Eqs.~\protect(\ref{alpha-comp}). To get
agreement with experiments we used  one fitting parameter $ J_0
\propto \chi $ and introduced an offset $J_{\rm offset}$ for the
current $J_x$ detected in the direction of light propagation. The
inset shows the experimental geometry. An additional inset in the
lower panel displays the sample and the radiation electric field
viewing from the source of radiation side. } \label{fig1}
\end{figure}

The experiments are carried out on (0001)-oriented wurtzite
$n$-GaN/Al$_{0.3}$Ga$_{0.7}$N heterojunctions grown by MOCVD on
C-plane sapphire substrates (for details of growth see
Ref.~\onlinecite{APL05}). The thickness of the AlGaN layers was
varied between 30~nm and 100~nm. An undoped 33~nm thick GaN buffer
layer grown under a pressure of 40~Pa at temperature
550$^{\circ}$C is followed by an undoped GaN layer ($\sim $
2.5~$\mu $m) grown under 40~Pa at 1025$^{\circ}$C; the undoped
Al$_{0.3}$Ga$_{0.7}$N barrier was grown under 6.7~Pa at
1035$^{\circ}$C. The mobility and  density in the two-dimensional
(2D) electron gas measured at room temperature are $\mu \approx
1200$~cm$^{2}$/Vs and $N_{s} \approx 10^{13}$~cm$^{-2}$,
respectively. To measure the photocurrent two  pairs of contacts
are centered at opposite sample edges with the connecting lines
along the axes $x \parallel [1 \bar{1} 00]$ and $y
\parallel [11\bar{2}0]$, see inset in Fig.~\ref{fig1}. The current
generated by the light in the unbiased samples is measured via the
voltage drop across a 50~$\Omega$ load resistor in a
closed-circuit configuration. The voltage is recorded with a
storage oscilloscope.

\begin{figure}[h]
\includegraphics[width=0.7\linewidth]{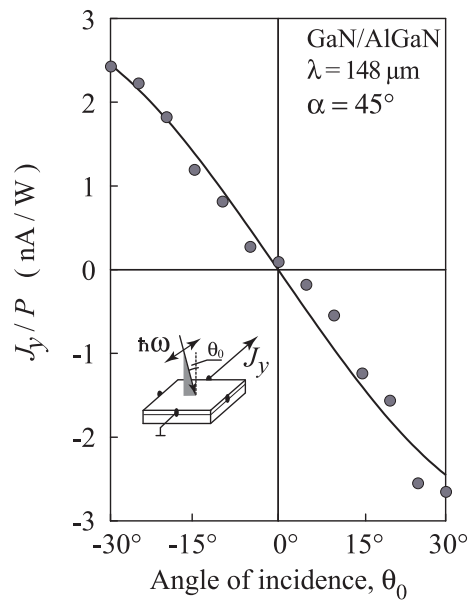}
\caption{Photocurrent as a function of angle of incidence
$\theta_0$ measured at room temperature in the transverse geometry
and azimuth angle $\alpha = 45^\circ$. Photocurrent is excited by
radiation with wavelength $\lambda$~=~148~$\mu$m and power $P
\approx 5$~kW. Full line is the fit to
Eqs.~\protect(\ref{alpha-comp}) with the same fitting parameter
$J_0$ as used in Fig.~\ref{fig1}. The inset shows the experimental
geometry. } \label{fig2}
\end{figure}

The photocurrents were induced by indirect intrasubband
(Drude-like) optical transitions in the lowest size-quantized
subband. The emission from a terahertz (THz) molecular laser
optically pumped by a TEA CO$_2$ laser\cite{6} is used for the
optical excitation. With NH$_{3}$, D$_{2}$O and CH$_{3}$F as an
active media we could obtain radiation pulses of duration $\simeq
$100~ns with wavelengths $\lambda =$ 77, 90.5, 148, 280, 385 and
496~$\mu $m and a power $P \simeq $ 5~kW. Radiation in both the
normal- and oblique-incidence geometries were applied with the
angle of incidence $\theta_0$ varying from $-30^\circ$ to
+30$^\circ$, $\theta_0 = 0$ corresponding to the normal incidence,
and laser beam lying in the $(xz)$ plane, see insets in
Figs.~\ref{fig1},~\ref{fig2} and~\ref{fig4}.

Optically pumped  molecular lasers emit linearly polarized
radiation with the polarization plane determined by that of the
pump light: the electric field vector of the THz radiation
$\bm{E}_{l}$ generated by the molecular laser can be either
parallel (like for the lines $\lambda =$ 90.5, 148 and 280~$\mu$m)
or orthogonal (lines 77, 385 and 496~$\mu$m) to the polarization
vector of the pump beam depending on the angular momentum
selection rules for the pump and the THz transitions in the active
media.\cite{6}

In the experiments the plane of polarization of the radiation
incident on the sample was rotated applying $\lambda/2$ plates
which enabled us to vary the azimuth angle $\alpha$ from $0^\circ$
to $180^\circ$ covering all possible orientations of the electric
field component in the ($xy$) plane. Hereafter the angle $\alpha =
0$ is chosen in such a way that the incident light polarization is
directed along the $y \parallel [11\bar{2}0]$ direction, see inset
in Fig.~\ref{fig1}(b).

To investigate the photogalvanic effects we also use elliptically
polarized light. In this case the polarization of the laser beam
is modified from linear to elliptical by means of crystal quartz
$\lambda/4$ plates. The counterclockwise rotation (viewing from
the laser side) of the optical axis of the quarter-wave plate by
the angle $\varphi_p$ results in the variation of the radiation
helicity as $P_{\rm circ} = - \sin{2 \varphi_p}$. Here the angle
$\varphi_p = 0$ is chosen for the position of the quarter-wave
plate optical axis coinciding with the incoming laser
polarization, in which case the linear polarization degree is
given by fourth harmonics of $\varphi_p$.

A series of measurements is carried out making use of the
frequency tunability and short pulse duration of the free electron
laser ``FELIX'' at FOM-Rijnhuizen in the Netherlands operated in
the spectral range between 70~$\mu$m and
120~$\mu$m.\cite{Knippels99p1578} The output pulses of light from
the FELIX were chosen to be $\approx$\,6~ps long, separated by
40~ns, in a train (or ``macropulse'') of 7~$\mu$s duration. The
macropulses had a repetition rate of 5~Hz.

\section{Experimental results}

Irradiating the (0001)-grown GaN/AlGaN heterojunction by polarized
light at oblique incidence, as sketched in the inset to
Fig.~\ref{fig1}(a), causes a photocurrent signal measured across a
contact pair. The width  of the photocurrent pulses is about
100~ns  which corresponds to the THz laser pulse duration. The
signal depends on the light polarization,  and all characteristic
polarization properties persist from 4.2 to 300~K. The
experimental data presented below are obtained at room
temperature. The effect is observed for all the wavelengths
applied (between 77~$\mu $m and 496~$\mu $m). We use two
geometries: the longitudinal geometry ($J_x$ in the
Fig.~\ref{fig1}), in which the photocurrent is measured in the
direction along in-plane component $\hat{\bm{e}}_{\parallel}$ and
the transverse geometry ($J_y$ in the Fig.~\ref{fig1}), where the
signal is detected in the direction normal to the light
propagation unit vector $\hat{\bm{e}}$  [see the inset in
Fig.~\ref{fig1}(a)].

Figure~\ref{fig1} shows the dependence of the photocurrent on the
azimuth angle $\alpha$ for both experimental geometries obtained
at the positive incidence angle $\theta_0 = 15^\circ$ and
$\hat{\bm{e}}_{\parallel} \parallel x$. The polarization
dependence of the current in the longitudinal geometry is well
fitted by $J_x = J_0 (1 - \cos{2 \alpha}) + J_{\rm offset}$ while
for the transverse geometry we have $J_y = - J_0 \sin{2 \alpha}$.
Note that the offset contribution $J_{\rm offset}$ is observed in
the longitudinal geometry only. Below, in Sec.~IV.B, we will
demonstrate that exactly these dependences follow from the theory.
As a function of the incidence angle the photocurrent changes sign
at $\theta_0 \approx 0$. This is demonstrated in Fig.~\ref{fig2}
which shows the dependence $J_y(\theta_0)$ for the fixed azimuth
angle $\alpha = 45^{\circ}$. A similar dependence is also detected
for the longitudinal current $J_x$. We note that, like the
polarization dependent contribution to the photocurrent, the
detected offset current in the longitudinal geometry $J_{\rm
offset}$ inverts its direction when the incidence angle changes
its sign.

\begin{figure}[h]
\includegraphics[width=0.9\linewidth]{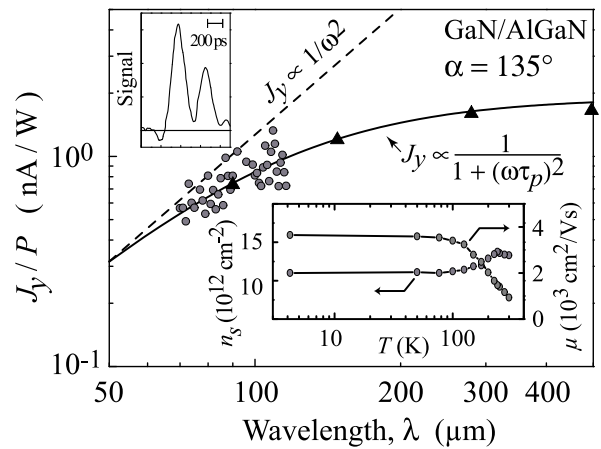}
\caption{Spectral dependence of the transverse photocurrent $J_y$
measured at room temperature at oblique incidence ($\theta_0 =
15^\circ$) for azimuth angle $\alpha = 135^\circ$. The data are
obtained with the free electron laser FELIX (dots) and with
molecular optically pumped laser (triangles). Full line  shows fit
to Eq.~\protect(\ref{abs}). The fit is obtained using $\tau_p$ as
an adjustable parameter and a scaling the whole dependence by the
ordinate. Dashed line shows $J_y \propto \omega^{-2}$ for
comparison. The  inset in the left upper corner shows the temporal
structure of the current in response to the radiation of FELIX.
The inset in the right corner shows the temperature dependences of
the carrier density and the  mobility. }
 \label{fig3}
\end{figure}

The wavelength dependence of the photocurrent obtained in
transverse geometry at azimuth angle $\alpha = 135^\circ$ and
angle of incidence $\theta_0 = 15^\circ$ is shown in
Fig.~\ref{fig3}. The data are measured both on FELIX (dots) and on
molecular laser (triangles). Fitting the data to the well known
spectral behavior of the Drude absorption\cite{Drude,Drude2}
\begin{equation} \label{abs}
J_y(\omega) \propto \eta(\omega) \propto {1 \over 1 +
(\omega\tau_p)^2}\:
\end{equation}
we obtain that the spectral behavior of the photocurrent can
reasonably be described by this equation (see full line in
Fig.~\ref{fig3}). For the fit we used $\tau_p$ as an adjustable
parameter and scaled the whole dependence by the ordinate.
Analyzing this dependence we obtained that the momentum relaxation
time controlling absorption is about 0.05~ps. This value is twice
shorter than the transport time ($\simeq 0.1$~ps) extracted from
the mobility measurements at room temperature. We attribute the
shortening of the momentum relaxation time by electron gas heating
due to absorption of the intense THz radiation. The reduction of
the mobility due to heating stems from  enhancement of  scattering
by phonons under increase of temperature. Using short  6~ps pulses
of  FELIX we observed that the response time is determined by the
time resolution of our set-up  but it is at most 100~ps or shorter
(see the inset in the left upper corner of Fig.~\ref{fig3}). This
fast response is typical for photogalvanics where the signal decay
time is expected to be of the order of the momentum relaxation
time\cite{SturmanFridkin,10,6} being in our samples at room
temperature of the order of  $0.1$~ps.

\begin{figure}[h]
\includegraphics[width=0.7\linewidth]{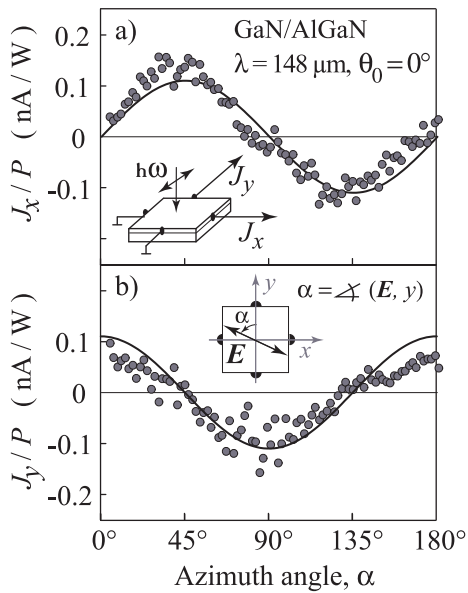}
\caption{ Photocurrent as a function of angle $\alpha$ measured at
normal incidence ($\theta_0 = 0$) in a)~$x$ and b)~$y$
crystallographic directions. Photocurrent is excited by linearly
polarized radiation with wavelength $\lambda$~=~148~$\mu$m and
power $P \approx 5$~kW.  Full lines are fits to
Eqs.~\protect(\ref{alpha_normal}). To get agreement with
experiments we used  one fitting parameter $J_1 \propto  \chi'$.
The inset shows the experimental geometry. An additional inset in
the lower panel displays the sample and the radiation electric
field viewing from the source of radiation side.}
 \label{fig4}
\end{figure}

In addition to the signal excited by the radiation at oblique
incident we also detected a small photocurrent at normal
incidence. The polarization behavior of this signal is  shown in
Fig.~\ref{fig4}. In this set-up photocurrent dependences can be
well fitted by $J_x = J_1  \sin{2 \alpha }$  and $J_y = J_1 \cos
{2 \alpha }$ (see also Sec.~IV.A). Rotating the sample around $z$
axis we proved that the current direction is solely determined by
the  orientation of radiation electric field relative to the
in-plane crystallographic axes. We note that this contribution is
about one order of magnitude smaller than that at oblique
incidence even at small angles $\theta_0 \approx 15^\circ$.

Besides investigations of photocurrent in response to linearly
polarized radiation we also performed measurements under
excitation with elliptically polarized light. Such experiments are
of particular interest because such radiation, in partially
circularly polarized light,  has previously been used for
investigation of the circular photogalvanic effect
(CPGE)\cite{10,APL05,6} which coexist with the LPGE yielding
beatings in $\varphi_p$-dependence. The CPGE is characterized by a
photocurrent whose direction is changed upon reversal of the
radiation helicity. It has been observed in GaN-based structures
demonstrating a substantial structural inversion asymmetry (SIA)
in this wide band gap
material.\cite{APL05,HeAPL2007,TangApl2007,ChoPRB2007} We have
shown that SIA in GaN-based heterostructures results in a spin
splitting of subbands in $\bm k$-space which was later confirmed
by magneto-transport
measurements.\cite{ChoApl2005,TangAPL2006,ThillosenAPL2006,SchmultPRB2006}
The Rashba spin-splitting due to SIA, which is not expected in
wide band semiconductors, in GaN/AlGaN heterostructures is caused
by a large piezoelectric effect\cite{CingolaniPRB2000} which
yields a strong electric field at the GaN/AlGaN interface and a
strong polarization induced doping effect.\cite{Litvinov}

\begin{figure*}
\includegraphics[width=0.75\linewidth]{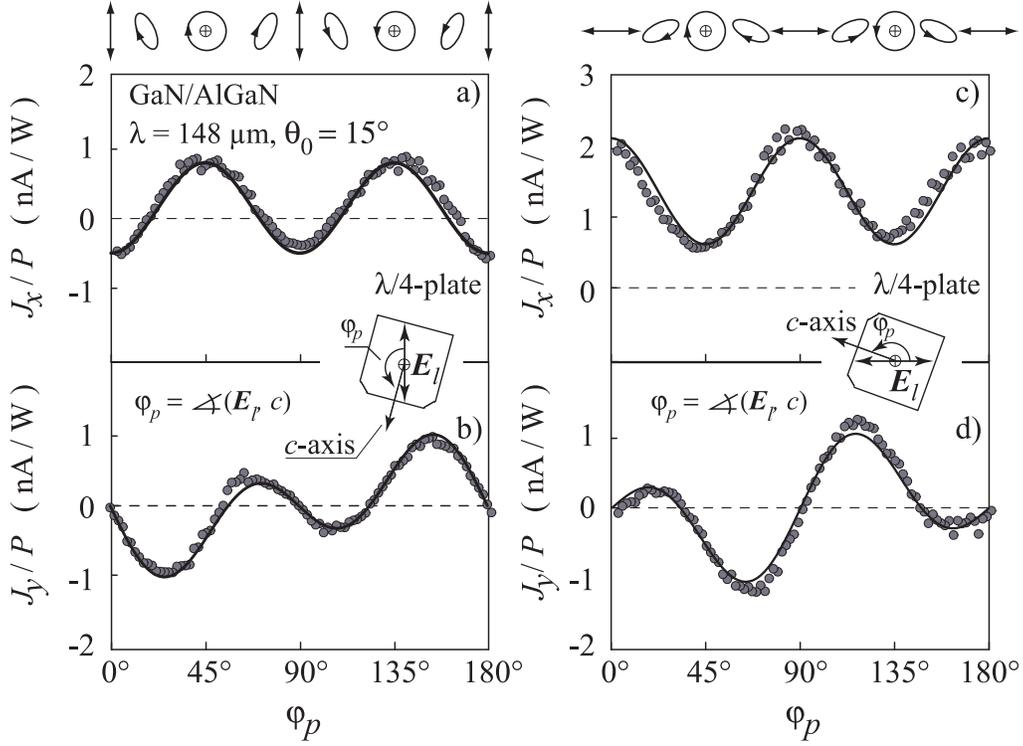}
\caption{Photocurrent as a function of angle $\varphi_p$ measured
at oblique incidence ($\theta_0 = 15^\circ$) in the longitudinal
[$J_x$ in a) and c)] and transverse [$J_y$ in b) and d)]
geometries at wavelength $\lambda$~=~148~$\mu$m and power $P
\approx 5$~kW. The ellipticity  of the radiation is varied by
passing linearly polarized laser radiation ($\bm{E}_l$)  through a
quarter-wave plate (see insets in the left and right panels). Left
panels show transverse and longitudinal photocurrents measured in
experimental set-up with $\bm{E}_{l}$ perpendicular to the
incidence plane  ($s$-polarization at $\varphi_p=0$, see inset in
the left panel). Photocurrents obtained for $\bm{E}_{l}$ parallel
to the incidence plane ($p$-polarization at $\varphi_p=0$, see
inset in the right panel) are shown in the right panel. Full lines
are fits of the photocurrent to Eqs.~\protect(\ref{phi-comp})
and~\protect(\ref{other_geometry_comp}) obtained correspondingly
in the experimental set-up  sketched in the inset in the left and
right panels. The fits are obtained using the same values of $J_0$
and $J_{\rm offset}$ as in the experiments with linearly polarized
radiation. The inset in the left panels shows the experimental
geometry. On top the polarization ellipses  corresponding to
various phase angles $\varphi_p$ are plotted viewing from the
source of radiation. }
 \label{fig5}
\end{figure*}

Figures~\ref{fig5}(a) and~\ref{fig5}(b) demonstrate the
dependences of the photocurrent on the  angle $\varphi_p$ for
positive angle $\theta_0$. In Fig.~\ref{fig5}(a) transverse and
longitudinal photocurrents are measured for the experimental
set-up with $\bm{E}_{l}$ perpendicular to the incidence plane
($s$-polarization at $\varphi_p=0$). We find that the polarization
dependences in this experiment are well fitted by $J_x = J_0 (1-
\cos{4 \varphi_p})/2 + J_{\rm offset}$  and for transverse
geometry by $J_y = - J_0  \sin{4 \varphi_p}\,/2 + J_2 \sin{2
\varphi_p}$, see Sec.~IV.B.2 for theoretical justification. We
note that we used for fitting the same values of $J_0$ and $J_{\rm
offset}$ as in experiments with linearly polarized radiation.

As we discussed above our laser can generate linearly polarized
radiation either oriented along or perpendicularly to the
polarization of the pump radiation. By that we change the position
of the electric field of transformed beam relative to the optical
axis of the quarter-wave plate, consequently changing the
orientation of the ellipse as well as radiation helicity of the
beam at the sample. In order to check how this transformation
influence  results we provided an additional experiment using
laser beam  with again  $\lambda = 148$~$\mu$m and
$\hat{\bm{e}}_\| \parallel x$ but for $\bm E_{l}$  parallel to the
incidence plane ($p$-polarization at $\varphi_p=0$).
Figures~\ref{fig5}(c),(d) show the data obtained in this geometry.
In contrast to the results presented in Figs.~\ref{fig5}(a),(b)
the photocurrent now shows another type of polarization
dependences: $J_x = J_0 (3 + \cos{4 \varphi_p})/2 + J_{\rm
offset}$  and for transverse geometry by $J_y =  J_0  \sin{4
\varphi_p}/2 + J_2  \sin{2 \varphi_p}$. The fact that contribution
given by $J_2$ did not change is not surprising because this term
is due to radiation helicity which does not change: $P_{\rm
circ}=-\sin {2 \varphi_p}$. The changes in contribution induced by
the degree of linear polarization are caused by the fact that the
orientation of corresponding ellipses at the sample at a given
$\varphi_p$ is different.

The photocurrent polarization properties revealed experimentally
both under normal and oblique incidence are explained in the next
Section where we propose a simple illustrative model to interpret
the nature of the photocurrent under study and derive general
phenomenological equations. The measured wavelength and
temperature dependences of the LPGE current are described in
Sec.~V where a microscopic theory is developed.

\section{Models, phenomenology and polarization dependences of the photocurrents}
\label{sec:Phenomen}

\begin{figure}[h]
\begin{center}
\includegraphics[width=0.9\linewidth]{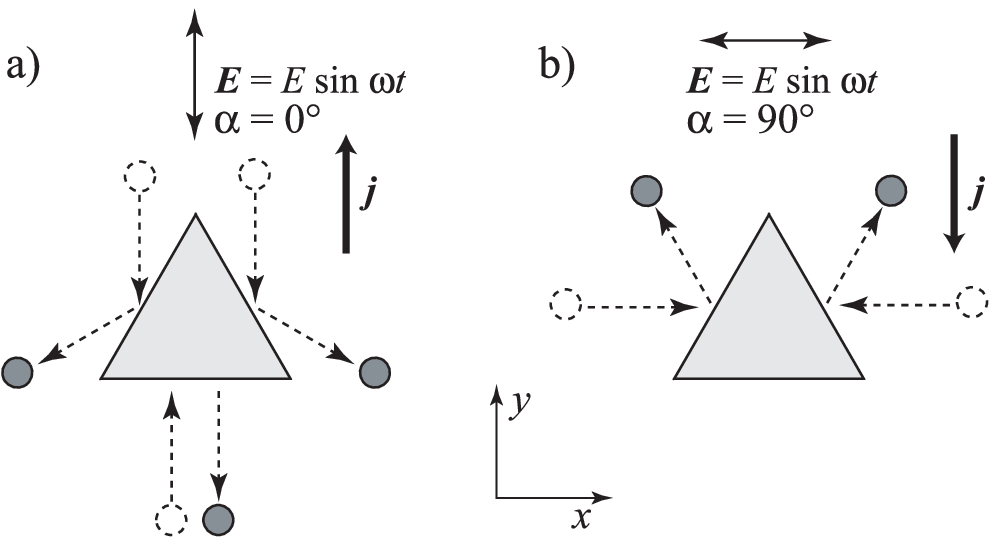}
\end{center}
\caption{Model of current generation due to  asymmetry of specular
elastic scattering by  wedges. The field  $\bm E$ results in
directed motion of carriers shown by dashed arrows. Due to
asymmetric scattering  a directed carrier  flow and, therefore,
electric current $\bm{j}$ are generated. a) and b) sketch two
relative orientations of wedges and the  electric field $\bm E$
together with resulting $dc$ current $\bm j$. }
\label{fig6}
\end{figure}

The appearance of the LPGE can be visualized by a simple
one-dimensional model of randomly distributed but identically
oriented wedges acting as asymmetric scattering
centers.\cite{SturmanFridkin,10,6} This model showing the
generation of a photogalvanic current is relevant for 2D GaN/AlGaN
structures excited by the radiation at normal incidence. In
Fig.~\ref{fig6} a wedge with a base oriented along $x$-direction
is depicted  which obviously does not possess a center of
inversion. In equilibrium the velocities of electrons are
isotropically distributed. Application of an external in-plane
alternating field ${\bm E}(t)= {\bm E} \sin \omega t$ adds an
oscillatory motion along the electric field to the random thermal
motion of the electrons. If the field points along the height of
the wedges (${\bm E}
\parallel y$, or  $\alpha = 0$), then the scattering results in a
carrier  flow in the $(-y)$ direction yielding an electric current
$j_y > 0$ shown by up-arrow in Fig.~\ref{fig6}(a). By that
electron fluxes along $x$ direction compensate each other and
$j_x$ is absent. A variation of the relative direction between the
electric field and the orientation of the wedges changes the
direction of the carrier flow resulting in a characteristic
polarization dependence. It may, e.g., reverse its direction, as
it is shown in Fig.~\ref{fig6}(b) for the field oriented along the
base of the wedges, or rotate by 90$^\circ$ in the case of $\bm E$
directed at 45$^\circ$ to $y$~axis.

In order to describe the observed polarization and angle of
incidence dependences, we derive here phenomenological equations
for the photocurrents in 2D GaN-based structures and give a model
for LPGE. The linear photogalvanic current density $\bm j$ is
phenomenologically described by the following
expression\cite{SturmanFridkin,10,6}
\begin{equation}
\label{LPGE} j^{LPGE}_{\lambda} = \sum_{\mu\nu}
\chi_{\lambda\mu\nu} (E_\mu E_\nu^* + E_\nu E_\mu^*)\:.
\end{equation}
Here $\bm E$ is the electric field amplitude of the light wave,
and $\bm \chi$ the third-rank tensor symmetric in the last two
indices. The index $\lambda$ enumerates two in-plane coordinates
$x$ and $y$, while $\mu$ and $\nu$ run over all three Cartesian
coordinates.

The LPGE is allowed only in piezoelectric crystal classes of
noncentrosymmetric media where nonzero components of the tensor
${\bm \chi}$ do exist. The point symmetry of wurtzite 2D
 electron systems confined in the [0001] direction
is C$_{3v}$ which differs from the $C_{6v}$ symmetry of the bulk
GaN due to absence of translation along the growth axis $z$, i.e.,
because of presence of interfaces. The coordinate frame  $x, y, z$
chosen in the experimental set-up means that the plane $(yz)
\parallel (1\bar{1}00)$ coincides  with  the mirror reflection planes
$\sigma_v$  contained in the C$_{3v}$ point group. In this point
group the tensor ${\bm \chi}$ has two linearly-independent
components for $\lambda \neq z$:
\begin{equation}
\chi \equiv \chi_{xxz} = \chi_{yyz}, \qquad \chi' \equiv
\chi_{xxy} = \chi_{yxx} = - \chi_{yyy}\:.
\end{equation}
This means that the phenomenological equation~(\ref{LPGE}) reduces
to
\begin{eqnarray} \label{lpgexy}
j^{LPGE}_x &=& \chi \{E_x E_z^*\} + \chi' \{E_x E_y^*\}, \\
j^{LPGE}_y &=& \chi \{E_y E_z^*\} + \chi' (|E_x|^2 -  |E_y|^2)\:,
\nonumber
\end{eqnarray}
where $ \{E_\mu E_\nu^*\} = E_\mu E_\nu^* + E_\nu E_\mu^*$.

\subsection{Normal incidence}
\label{Sec:normal}

In our experiments, a linear photogalvanic current is observed at
normal incidence, see Fig.~\ref{fig4}. It follows from
Eqs.~\eqref{lpgexy}  that while at normal incidence  the
photocurrent proportional to the constant $\chi$ vanishes, because
$E_z = 0$, the contribution determined by the constant $\chi'$ is
nonzero.  The observation of this photocurrent is an important
result demonstrating substantial difference in symmetry between
wurtzite 2D GaN(0001)-based structures with $C_{3v}$ symmetry and
zinc-blende GaAs- and InAs-based heterostructures having $C_{2v}$
or $D_{2d}$ symmetry, where the LPGE is forbidden at normal light
incidence.

The polarization dependences for normal incidence  follow from
Eqs.~\eqref{lpgexy}  and are given by
\begin{eqnarray} \label{alpha_normal}
&&j_{x} =  \chi' (t_0 E_0)^2  \sin{2 \alpha} \:, \\
&&j_{y} =  \chi' (t_0 E_0)^2   \cos{2\alpha} \:, \nonumber
\end{eqnarray}
where $t_0=2/(n_\omega+1)$ is the amplitude transmission
coefficient for normal incidence, $n_{\omega}$ is the refractive
index of the medium (for GaN $n_{\omega} = 2.3$). Both
polarization dependences well describe our experimental data with
one fitting parameter $\chi'$ (see solid lines in
Fig.~\ref{fig4}). This agreement clearly demonstrates the
generation of LPGE current at normal incidence. We emphasize that
the current direction depends on the orientation of the
polarization plane in respect to the crystallographic axes $x$ and
$y$. In the case that contacts are arbitrary oriented relative to
the crystallographic directions Eqs.~\eqref{alpha_normal} hold but
the phase shift appears. This is considered in Appendix~A.

\subsection{Oblique incidence}

\subsubsection{Linear polarization}

Linear photogalvanic effect is also observed at oblique incidence.
Moreover, comparison of Figs.~\ref{fig1} and~\ref{fig4} shows that
even at a small incidence angle $\theta_0$ the photocurrent  is
substantially (by the order of magnitude) larger than that at
normal incidence. While at normal incidence the photogalvanic
current is solely described by the constant $\chi'$ at oblique
incidence another term in Eqs.~\eqref{lpgexy} shows up. This term
is determined by the second linearly independent constant $\chi$.
The polarization dependence of this contribution in the geometry
relevant to the experiment depicted in Fig.~\ref{fig1}, where the
incidence plane is chosen to be the plane ($xz$) and the angle
$\alpha$ is counted from $y$, is given by
\begin{eqnarray} \label{alpha-comp}
&&j_x(\alpha) =  \chi E_0^2 t_p^2 \cos{\theta} \sin{\theta}  (1 - \cos{2\alpha})\:, \\
&&j_y(\alpha) =  - \chi E_0^2 t_p t_s \sin{\theta} \sin{2\alpha}
\:. \nonumber
\end{eqnarray}
Here $\theta$ is the refraction angle related to the incidence
angle $\theta_0$ by $\sin{\theta} = \sin{\theta_0}/n_\omega$, and
$t_s$ and $t_p$ are the Fresnel amplitude transmission
coefficients from vacuum to the structure for the $s$-  and
$p$-polarized light, respectively.\cite{Born_Wolf} These functions
are shown in Fig.~\ref{fig1} by solid lines.  To get agreement
with experiments we used  the above equations with one fitting
parameter $\chi$ and introduced an offset for the current $J_x$
detected in the direction of the light propagation. Registration
of these characteristic polarization dependences proves
observation of LPGE at oblique incidence.

Comparing the results presented in Figs.~\ref{fig1} and~\ref{fig4}
we can estimate of the ratio of different components of the tensor
$\bm \chi$ for the studied structures. At $\theta \approx
0.1$~rad, $t_s \approx t_p \approx t_0$, therefore from the ratio
of the amplitudes in Figs.~\ref{fig1} and~\ref{fig4} equal to
$|\chi'/\chi \, \theta|$ we get $|\chi'/\chi| \approx 10^{-2}$.
This hierarchy of components of the third rank tensor
$\chi_{\lambda\mu\nu}$ in the systems of $C_{3v}$ symmetry is
expectable,\cite{SturmanFridkin} however we demonstrate that the
$\chi'$-related effect is observable. While the effect described
by the constant $\chi$ also exists in the bulk GaN as well as in
the $C_{\infty v}$ group the photocurrent proportional to $\chi'$
arises only due to the reduced symmetry $C_{3v}$ of the system,
i.e. due to size quantization. Therefore the latter contribution
should increase at narrowing of 2D layer.

Equations~(\ref{alpha-comp}) show that the photogalvanic current
should follow the dependence $t_s t_p \sin{\theta}$, in
particular, reverse its direction upon inversion the angle of
incidence. This behavior is indeed observed. Figure~\ref{fig2}
shows LPGE current dependence on the incidence angle, $\theta_0$
obtained for $J_y$ at a fixed linear polarization of radiation
($\alpha = 45^\circ$). We note that the offset photocurrent
observed in the longitudinal direction only is always directed
antiparallel to $\hat{\bm e}_\|$. We ascribe the offset to the
photon drag current which is linearly coupled to the photon
momentum. This effect is out of scope of the present paper.

In our experiments we also probe LPGE for the light directed along
other crystallographic directions. Investigating transverse and
longitudinal currents being perpendicular and parallel to the
incidence plane, respectively, we observed  that all polarization
dependences are described by the same constant $\chi$
independently of the in-plane direction  of the light propagation.
The reason of this fact is that the current contribution
proportional to $\chi$ is due to cone asymmetry being  axial
isotropic in the plane of our structures. Thus this photocurrent
is totally determined by the light polarization state and
propagation direction,
 and it is independent of the incidence plane orientation relative
to the in-plane crystallographic axes. For linearly polarized
light the photocurrent can be presented in the invariant form as
${\bm j} = 2 \chi {\bm E}_\parallel E_z$ where ${\bm E}_\parallel$
is the projection of the light polarization vector $\bm E$ onto
the structure plane. It is seen that the LPGE current is always
directed along ${\bm E}_\parallel$. By introducing the azimuth
angle $\beta$ so that $\beta =0 $ corresponds to the polarization
perpendicular to the incidence plane ($s$-polarization) we obtain
for transverse (${\bm j}_\perp$)  and longitudinal (${\bm
j}_\parallel$) currents
\begin{eqnarray} \label{alpha-depend}
&&{\bm j}_\perp =  \chi E_0^2 t_p t_s \sin{2\beta} \, (\hat{\bm e}_\| \times \hat{\bm z}) \:, \\
&&{\bm j}_\parallel =  \chi E_0^2 t_p^2 \cos{\theta} \, (1 -
\cos{2\beta})\hat{\bm e}_\| \:, \nonumber
\end{eqnarray}
where $\hat{\bm z}$ is the unit vector along the normal.

\subsubsection{Elliptical polarization}

Illumination of the structure by elliptically polarized radiation
also results in generation of the LPGE current which is determined
by the linear polarization degree of the light. However,
elliptically polarized radiation is also characterized by the
nonzero helicity. This results in the additional effect solely
determined by the degree of circular polarization, the circular
photogalvanic effect.\cite{SturmanFridkin,10,6} This effect is
phenomenologically described by
\begin{equation}
\label{CPGE_def} j^{CPGE}_{\lambda} = \sum_{\mu}
\gamma_{\lambda\mu} \, {\rm i} ({\bm E} \times {\bm E}^*)_\mu =
 E_0^2 t_p t_s P_{circ}({\bm \gamma} \hat{\bm{e}})_\lambda   \:,
\end{equation}
where $\hat{\bm{e}}$ is the light propagation unit vector and $\bm
\gamma$  the second-rank pseudotensor. In the systems of $C_{3v}$
symmetry the tensor $\bm \gamma$ has one linearly-independent
component, namely $\gamma_{xy} = - \gamma_{yx} \equiv \gamma$.
Thus, the CPGE current flows always perpendicularly to the
incidence plane.

In our experimental set-up a quarter-wave plate is rotated by the
angle $\varphi_p$ between the laser light polarization vector,
$\bm{E}_l$, and the principal axis of the polarizer (see inset to
Fig.~\ref{fig5}). The total photogalvanic current is given by both
second and fourth harmonics of this angle. While radiation
handedness is solely determined by the angle $\varphi_p$ and does
not depend on the relative orientation between $\bm{E}_l$ and the
plane of incidence the orientation of the ellipse is substantially
different.
In particular, for $\bm{E}_l$ perpendicular to the incidence plane
at $\varphi_p=0$ ($s$-polarization) and for $\bm{E}_l$ parallel to
the incidence plane at $\varphi_p=0$ ($p$-polarization) the
orientation of corresponding ellipses on the sample differs by
90$^\circ$.
Therefore in the former case light is never $p$-polarized and in
the latter case $s$-polarization can not be achieved, which is
quite different from the geometry of half-wave plate where all
states of linear polarization can be obtained. Thus we derive two
sets of equations describing experimental set-up of
Figs.~\ref{fig5}(a),(b) and~\ref{fig5}(c),(d).

In the first geometry ($s$-polarization at $\varphi_p=0$)  from
Eqs.~\eqref{lpgexy} and \eqref{CPGE_def} we get
\begin{eqnarray} \label{phi-comp}
&&j_x({\varphi}_p) = {\chi \over 2} E_0^2 t_p^2 \cos{\theta} \sin{\theta} \:  (1- \cos{4\varphi_p})   \:, \\
&&j_y({\varphi}_p) = E_0^2 t_p t_s \sin{\theta} \: \left( -{\chi
\over 2} \sin{4\varphi_p} + \gamma \sin{2\varphi_p} \right) \:,
\nonumber
\end{eqnarray}
and for the latter geometry ($p$-polarization at $\varphi_p=0$)
\begin{eqnarray}
    \label{other_geometry_comp}
&&j_x({\varphi}_p) = {\chi \over 2} E_0^2 t_p^2 \cos{\theta} \sin{\theta} \:  (3 + \cos{4\varphi}_p) \:, \\
&&j_y({\varphi}_p) = E_0^2 t_p t_s \sin{\theta} \: \left(  {\chi
\over 2} \sin{4\varphi_p} + \gamma \sin{2\varphi_p}  \right) \:.
\nonumber
\end{eqnarray}
Here we again took into account that in our samples $|\chi'| \ll
|\chi|$. Fits of experiment with the constant $\chi$ independently
obtained from experiments with half-wave plates and $\gamma$ as a
single adjustable parameter being the same for all curves yields a
good agreement with the experiment. Here again for the
longitudinal photocurrent the offset due to photon drag effect
shows up and has the same value as that in the experiments with
half-wave plates.

\section{Microscopic theory}
\label{sec:Micro_theory}

In order to describe the measured dependences of the photocurrent
on wavelength and temperature we have developed a microscopic
theory of LPGE. First we consider the contribution proportional to
the coefficient $\chi$ in Eqs.~(4). This contribution does exist
not only in systems of the C$_{3v}$ or C$_{6v}$ symmetry but is
also allowed by the uniaxial point group C$_{\infty v}$.
Therefore, we can calculate the $\chi$ contribution to $j^{\rm
LPGE}_{x,y}$ in the uniaxial approximation. Then we turn to
calculation of the contribution proportional to $\chi'$ in
Eqs.~(4) and surviving under normal incidence due to the reduced
C$_{3v}$ symmetry of the system. We note, that so far the LPGE in
2D systems has been treated theoretically for direct intersubband
transitions only.\cite{Magarill_Entin,book_Ivchenko_Pikus} The
mechanism of the LPGE under Drude absorption differs strongly
because scattering is unavoidably present in the intrasubband
absorption process.

\subsection{Microscopic model for oblique incidence}
\label{sec:Oblique}

Microscopically the LPGE current consists of the shift and
ballistic contributions which in general can be comparable in
order of magnitude. We start from the analysis of the shift
contribution and then turn to the ballistic contribution.

\subsubsection{Shift contribution}
\label{sec:Shift}

The shift contribution arises due to  coordinate shifts of the
free-carrier wave-packet center-of-mass by microscopic lengths in
quantum transitions. The corresponding current is a product of the
elementary charge $e$, the transition probability rate $W_{{\bm
k}' {\bm k}}$ and the shift vector ${\bm R}_{{\bm k}' {\bm k}}$
or, explicitly,
\begin{equation}
\label{j_shift_def} {\bm j}_{\rm LPGE}^{({\rm shift})} = 2 e
\sum_{{\bm k},{\bm k}'} W_{{\bm k}' {\bm k}} \, {\bm R}_{{\bm k}'
{\bm k}}\:,
\end{equation}
where the factor ``two'' accounts for spin degeneracy. The
elementary shift is related to the transition matrix element
$M_{{\bm k}' {\bm k}}$ by\cite{BelIvchStur}
\begin{eqnarray}
\label{R_def} {\bm R}_{{\bm k}' {\bm k}} = - {{\rm Im}[ M^*_{{\bm
k}' {\bm k}} (\nabla_{\bm k} + \nabla_{{\bm k}'}) M_{{\bm k}' {\bm
k}}] \over \left| M_{{\bm k}' {\bm k}} \right|^2}  \\ = -
(\nabla_{\bm k} + \nabla_{{\bm k}'}) {\rm arg}(M_{{\bm k}' {\bm
k}})\:,  \nonumber
\end{eqnarray}
where ${\rm arg}(u)$ is the phase of the complex number $u$. The
above two equations are general and valid for the shift
photocurrent calculation in any frequency
range.\cite{SturmanFridkin,10} Here, for the first time, we apply
them to consider the LPGE under the Drude absorption. Note that
Eq.~(12) was applied in the theory of anomalous Hall
effect\cite{sinitsyn,sinitsyn2} where the coordinate shift at
scattering is called a side-jump.

\begin{figure}[h]
\includegraphics[width=0.7\linewidth]{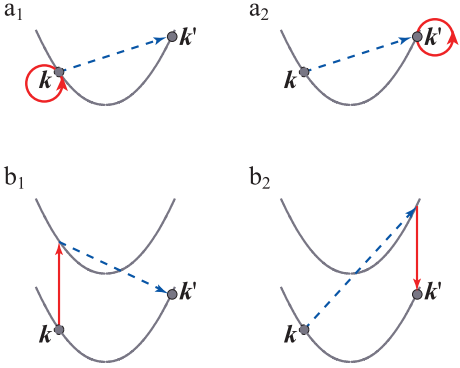}
\caption{The quantum transitions responsible  for the shift
contribution to the photocurrent. Note that Drude absorption under
oblique light incidence is caused by a$_1$ and a$_2$ processes
only.}
 \label{DrudeAbs}
\end{figure}

The main contribution to the matrix element for indirect
intrasubband optical transitions,
\begin{equation} \label{M_e_par}
M_{1{\bm k}', 1 {\bm k}}^{\parallel} = U_{1{\bm k}',1 {\bm k}}
\left( {V_{{\bm k}'} \over E_{{\bm k}'} - E_{\bm k}}  - {V_{\bm k}
\over \hbar \omega}\right)\:,
\end{equation}
comes from two-quantum processes a$_1$ and a$_2$ with intermediate
virtual states in the same $e1$ subband, see Fig.~\ref{DrudeAbs}.
Here the 2D electron kinetic energy $E_{\bm k} = \hbar^2 k^2/ 2
m$, $m$ is the electron in-plane effective mass, $V_{\bm k}$ is
the matrix element of electron-light interaction
\[
V_{\bm k} = - {e \hbar  A \over m c}\ {\bm e} \cdot {\bm k} \:,
\]
$\omega$, $A$ and ${\bm e}$ are the frequency, amplitude and unit
polarization vector of the light wave, $U_{1{\bm k}',1 {\bm k}}$
is the matrix element of intrasubband elastic scattering $e1, {\bm
k} \to e1, {\bm k}'$, the superscript $\parallel$ indicates that
the process is allowed for the polarization ${\bm e}$ containing
an in-plane component ${\bm e}_{\parallel}$. In the following, for
simplicity, we ignore the wave vector dependence of $U_{1{\bm
k}',1 {\bm k}}$ and replace it by a constant $U_{11}$. The
probability rate $W_{{\bm k}' {\bm k}}$ is given by Fermi's golden
rule and expressed via the squared modulus of the transition
matrix element and $\delta$-function describing the energy
conservation
\begin{equation} \label{energy}
E_{{\bm k}'} - E_{\bm k} = \hbar \omega\:.
\end{equation}

The processes a$_1$ and a$_2$ in isolation from other processes
make no contribution to ${\bm j}_{\rm LPGE}^{({\rm shift})}$,
because $M_{1{\bm k}',1 {\bm k}}^{\parallel}$ is independent of
$e_z$ while the $\chi$-related photocurrent in the
phenomenological equations~(\ref{lpgexy}) is proportional to
$e_z$, or because the microscopic expression in the square
brackets in Eq.~(\ref{R_def}) is real and hence the elementary
shift is absent. The shift contribution becomes nonzero if two
other indirect processes, b$_1$ and b$_2$ in Fig.~\ref{DrudeAbs},
are taken into account. The virtual states in these processes lie
in the second size-quantized subband $e2$. The corresponding
matrix element reads as
\begin{equation} \label{M_e_z}
M_{1{\bm k}', 1 {\bm k}}^{\perp} = - {\rm i} e_z {eA \over \hbar
c} z_{21} \Delta_{21} \left( {U_{12} \over \Delta_{21} - \hbar
\omega}  - {U_{21} \over \Delta_{21} +  E_{{\bm k}'} - E_{\bm k}}
\right).
\end{equation}
Here $U_{21} = U_{12}$ is the intersubband scattering matrix
element, $\Delta_{21}$ is the energy spacing between the $e2$ and
$e1$ subbands, and $z_{21}$ is the intersubband matrix element of
the coordinate $z$. The superscript $\perp$ indicates that the
transitions (\ref{M_e_z}) are allowed in the polarization
perpendicular to the interface plane.

Substituting $M_{{\bm k}' {\bm k}} = M_{{\bm k}' {\bm
k}}^{\parallel} + M_{{\bm k}' {\bm k}}^{\perp}$ into
Eq.~(\ref{R_def}), taking then into account the energy 
conservation equation (\ref{energy}) and assuming $\hbar \omega
\ll \Delta_{21}$ we obtain for the elementary shift
\begin{eqnarray} \label{shiftkk'}
&& \left| M_{{\bm k}' {\bm k}} \right|^2 {\bm R}_{{\bm k}' {\bm k}}  \\
&& = - z_{21 }\left( \frac{e\hbar A}{ m c} \right)^2 \frac{U_{11}
U_{21}}{\Delta_{21} \hbar \omega} e_z [{\bm e} \cdot (3 {\bm k}' -
{\bm k})]\ ({\bm k}' - {\bm k})\:. \nonumber
\end{eqnarray}
Note that the shift is different from zero only due to the wave
vector dependence of denominators $E_{{\bm k}'} - E_{\bm k}$ and
$\Delta_{21} +  E_{{\bm k}'} - E_{\bm k}$ in the matrix
elements~(\ref{M_e_par}) and (\ref{M_e_z}).

The result for the shift photocurrent can be presented in the form
\begin{equation} \label{j_shift_result}
{\bm j}_{\rm LPGE}^{({\rm shift})} = - e \xi  z_{21} {\bm
e}_{\parallel} e_z \frac{\eta_\parallel I}{\Delta_{21}}\ \frac{ F
\left[ 4 E + 3 \hbar \omega \right]}{F \left[ 2 E +  \hbar \omega
\right]}\:.
\end{equation}
Here the factor $\xi$ is given by\cite{ST_CPGE_orbital}
$$\xi = {\langle U_{11}U_{21} \rangle \over \langle U^2_{11} \rangle} ,$$
where the angular brackets mean averaging over the distribution of
static scatterers along $z$, the functional $F[\phi(E)]$ turns any
function $\phi(E)$ into a real number as follows
\[
F[\phi(E)] = \int\limits_0^\infty [f(E) - f(E + \hbar \omega)]
\phi(E) dE \left/ \int\limits_0^\infty f(E) dE \right.\:,
\]
$f(E)$ is the Fermi-Dirac distribution function, $\eta_\parallel$
is the absorbance under normal incidence ($\theta =0$) given
by\cite{ST_spin_orient}
\begin{equation} \label{Drude_absorbance}
\eta_\parallel = \frac{e^2}{\hbar c}\ {2 \pi N_s \over n_{\omega}
m \omega^3 \tau_p}\ F \left[ 2E + \hbar \omega \right]\:,
\end{equation}
the momentum relaxation time $\tau_p = \hbar^3/ (m N_d \langle
U^2_{11} \rangle)$, $N_s$ and $N_d$ are the 2D concentrations of
electrons and scatterers.

Note that allowance for nonzero value of $\xi$ and, hence, of the
LPGE current, presupposes the uniaxial asymmetry of the
structure.\cite{ST_CPGE_orbital} This corresponds to phenomenology
of the LPGE which requires absence of an inversion center in the
system.  A value of $\xi$ is nonzero due to an asymmetrical shape
of the electron envelope functions in the $e1$ and $e2$ subbands
caused by the build-in electric field. An inhomogeneous
asymmetrical distribution of scatterers along the $z$ axis can
also contribute to $\xi$.

\subsubsection{Ballistic contribution}
\label{sec:Ballistic}

\begin{figure*}
\includegraphics[width=0.9\linewidth]{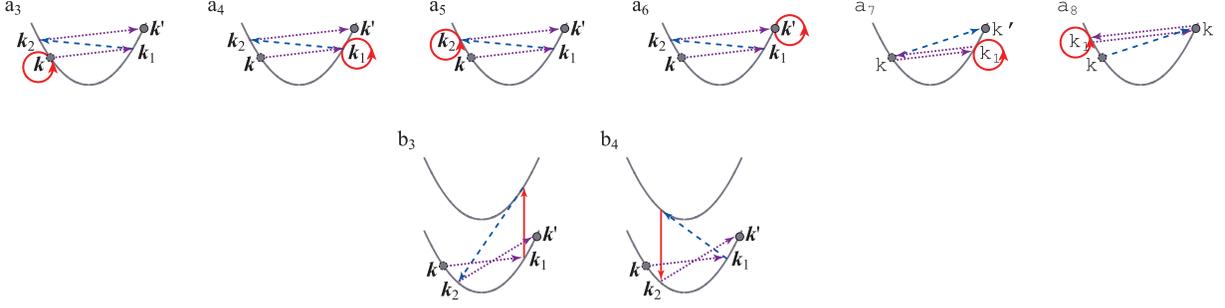}
\caption{Optical two-impurity-mediated processes included into
consideration in order to obtain nonvanishing antisymmetric part
of the scattering rate  $W^{a}_{{\bm k}'{\bm k}}$ in
Eq.~(\protect\ref{W}). For the processes a$_{3 \ldots6}$, the wave
vectors satisfy the relation $\bm{k} + \bm{k}_2 = \bm{k}' +
\bm{k}_1$.}
 \label{Ballistic}
\end{figure*}

In order to obtain a nonzero ballistic photocurrent it is
necessary to go beyond the approximation of single-impurity
scattering and take into account simultaneous scattering by two
impurities in one process. Such optical-absorption processes are
schematically shown in Fig.~\ref{Ballistic}. Let us introduce the
antisymmetrical part, $W^{a}_{{\bm k}' {\bm k}}$, of the ${\bm k}
\to {\bm k}'$ transition rate $W_{{\bm k}' {\bm k}}$ which changes
sign under simultaneous inversion of the wave vectors ${\bm k}$
and ${\bm k}'$. It can be expressed in terms of Fermi's golden
rule
\begin{equation}
\label{W} W^{a}_{ {\bm k}' {\bm k} } = {2 \pi \over \hbar} \left|
M_{{\bm k}' {\bm k}} \right|_a^2 \, [f(E_{\bm k}) - f(E_{{\bm
k}'})] \, \delta (E_{{\bm k}'} - E_{\bm k} - \hbar \omega)\:,
\end{equation}
where $\left| M_{{\bm k}' {\bm k}} \right|_a^2$ is the
antisymmetric part of the squared modulus of the optical
two-impurity-mediated matrix element $M_{{\bm k}' {\bm k}}$. Then
the ballistic contribution to the LPGE current is given by
\begin{equation}
\label{j_ball_def} {\bm j}_{\rm LPGE}^{({\rm ball})} = 2 e
\sum_{{\bm k},{\bm k}'} W^{a}_{{\bm k}'{\bm k}} \, \tau_p \ ({\bm
v}_{{\bm k}'} - {\bm v}_{\bm k})\:,
\end{equation}
where ${\bm v}_{\bm k} = \hbar {\bm k}/m$ is the electron
velocity.

Here it is worth to discuss the dependence of the ballistic and
shift contributions on the concentration of elastic scatterers
$N_d$. For two-impurity scattering the probability rate
$W^{a}_{{\bm k}'{\bm k}} \propto N_d^2$. In contrast, the
single-impurity scattering rate $W_{{\bm k}'{\bm k}}$ in Eq.~(11)
linearly depends on $N_d$. However,  the ballistic current is
controlled by the momentum relaxation time, see the factor
$\tau_p$ in the right-hand side of Eq.~(20). Since $\tau_p \propto
N^{-1}_d$, the ballistic and shift currents are both proportional
to $N_d$. In this sense, there is an analogy between a pair of the
two contributions to the photocurrent and a pair of the side-jump
and skew-scattering currents in the anomalous Hall
effect.\cite{sinitsyn2}

In order to get the photocurrent proportional to both ${\bm
e}_\parallel$ and $e_z$, we extract from $\left| M_{{\bm k}' {\bm
k}} \right|^2$ the interference term proportional to $ {\rm Re}
\left\{ M^{\parallel} M^{\perp*} \right\}$, where $ M^{\parallel}$
and $M^{\perp}$ are the contributions to the matrix element of the
indirect optical transition proportional to $e_{x, y}$ and $e_z$,
respectively. To the lowest order in $\hbar \omega / \Delta_{21}
\ll 1$, one should include into consideration six two-impurity
scattering processes a$_3 \ldots$a$_8$ contributing to
$M^{\parallel}$ and two processes b$_3$, b$_4$ for $M^{\perp}$
depicted in Fig.~\ref{Ballistic}.

For example, below we present the contributions to $M^{\parallel}$
from the processes a$_5$ and a$_6$,
\begin{eqnarray}
\label{M_e_par_4order}
&&M_{a_5}^{\parallel}  = N_d \times
\nonumber \\
\mbox{} \\
&& \sum_{\bm{k}_1 \bm{k}_2} { U'_{11} \, V_{{\bm k}_2} \, U_{11}
\, U'_{11} \  \delta_{\bm{k} + \bm{k}_2, \bm{k}' + \bm{k}_1} \over
(E_{\bm{k}_2} - E_{\bm{k}} - \hbar \omega - {\rm i}0)
(E_{\bm{k}_2} - E_{\bm{k}} - {\rm i}0) (E_{\bm{k}_1} - E_{\bm{k}}
- {\rm i}0)}, \nonumber
\end{eqnarray}
\begin{eqnarray}
&&M_{a_6}^{\parallel}  =  N_d \times
\nonumber \\
&& \sum_{\bm{k}_1 \bm{k}_2} { V_{{\bm k}'} \, U'_{11} \, U_{11} \,
U'_{11} \  \delta_{\bm{k} + \bm{k}_2, \bm{k}' + \bm{k}_1} \over
(E_{\bm{k}'} - E_{\bm{k}}  - {\rm i}0) (E_{\bm{k}_2} - E_{\bm{k}}
- {\rm i}0) (E_{\bm{k}_1} - E_{\bm{k}} - {\rm i}0)}, \nonumber
\end{eqnarray}
where the prime means the scattering (shown by dotted lines in
Fig.~\ref{Ballistic}) by a defect different from that which is
related to the matrix element $U_{11}$ in these equations and to
$U_{21}$ in Eq.~(\ref{M_e_z}) (shown by dashed lines). The
Kronecker symbols are obtained after averaging the matrix element
over the in-plane distribution of the second scatterer. Note that
averaging over its distribution in the $z$ direction results in
replacement of the square $(U_{11}^{\prime})^2$ by $\langle
U_{11}^2 \rangle$. The energy denominators are rewritten by using
the identity
\[
{1 \over x - {\rm i}0} = PV {1 \over x} + {\rm i} \pi \delta(x)\:,
\]
and, among three such terms in each contribution
$M^{\parallel}_{a_j}$ $(j = 3...8)$, we retain the
$\delta$-function in one of them and take the Cauchy principal
values for two others. This additional $\delta$-function, together
with the energy conservation law~(\ref{energy}) and the relation
$\bm{k} + \bm{k}_2 = \bm{k}' + \bm{k}_1$, fixes absolute values of
the wave vectors ${\bm k}, {\bm k}', {\bm k}_1$ and ${\bm k}_2$.
The integration over the azimuthal angle of the vector ${\bm k}_1$
either ${\bm k}_2$ is performed by using the principal value
integral
\[
PV \int\limits_0^{2\pi} {d \varphi \over 2\pi} {1 \over a - b
\cos{\varphi}} = { \Theta(a^2 - b^2)\ {\rm sign}\,a\ \over
\sqrt{a^2 - b^2}}\:,
\]
where $\Theta(x)$ is the Heaviside step-function. The calculation
shows that, among eight processes presented in
Fig.~\ref{Ballistic}, only two, namely, the processes a$_5$ and
a$_6$, lead to nonzero ballistic current. The final result for
this current reads
\begin{equation} \label{j_ball_result}
{\bm j}_{\rm LPGE}^{({\rm ball})} = {\bm e}_\parallel e_z
{\eta_\parallel I \over \Delta_{21}} \xi e z_{21} { F [ L(\hbar
\omega / E)] \over F [ E/\hbar \omega + 1/2 ]}\:,
\end{equation}
where
\[
L(x) = {x \over 2 \pi} \int\limits_{\varphi_0(x)}^{2\pi} {d
\varphi \over \sqrt{(x-2\sqrt{1+x} \cos{\varphi})^2 - 4}}\:,
\]
\[
\varphi_0(x) = \Theta(8-x) \arccos{\left( {x-2 \over
2\sqrt{1+x}}\right)}.
\]
The function $L(x)$ goes to zero as $x/4$ at $x \to 0$, and tends
to 1 at $x \to \infty$. A sum of the shift and ballistic
contributions given by Eqs.~(17) and~(20) represents the
photocurrent phenomenologically described by the coefficient
$\chi$ in Eqs.~(4).

\subsection{Microscopic model for normal incidence}
\label{sec:Normal}

An ideal GaN/AlN interface has the point-group symmetry C$_{3v}$
which is a subgroup of the C$^4_{6v}$ space group of a bulk
wurtzite crystal. The group C$_{6v}^4$ is nonsymmorphic and
contains, in addition to $C_3$ and $\sigma_v$, the elements
$\{C_6| {\bm \tau} \}, \{\sigma_d| {\bm \tau} \}$, where ${\bm
\tau}$ is the fractional lattice translation (0,0,$c/2$) and $c$
is the lattice constant along the principal axis $z$. In the 2D
structures, the screw-axis and glide-plane operations shift the
interfaces along $z$ not bringing the system into coincidence with
itself and, as a result, the point-group symmetry reduces to
C$_{3v}$.

For the C$_{6v}$ symmetry of bulk wurtzite structure,
characteristic of III-nitrides, the nonvanishing third order
tensor components $\chi_{\lambda \mu \nu}$ symmetrical in $\mu$
and $\nu$ are $\chi_{zzz}$, $\chi_{zxx} = \chi_{zyy}$,
$\chi_{xxz}= \chi_{xzx} = \chi_{yyz} = \chi_{yzy} \equiv \chi$.
Therefore, forbidden are such bulk effects as the second-harmonic
generation for excitation along $z$, a piezoelectric field induced
in the $(x,y)$ plane under the deformation in this plane and the
photocurrent in the $(x,y)$ plane for the light propagating along
$z$. The reduction from C$_{6v}$ to C$_{3v}$ removes these
restrictions and four new nonzero components appear, $\chi_{xxy} =
\chi_{xyx} = \chi_{yxx} = - \chi_{yyy} \equiv \chi'$, see
Eqs.~(\ref{lpgexy}). Note that it is the interface that is
exclusively responsible for the appearance of new components.

Here we propose a microscopic model of the LPGE under normal
incidence. In this geometry virtual transitions via other
size-quantized electronic subbands are forbidden, therefore we
should consider the light absorption with intermediate states
either in the same $e1$ electronic subband or in the valence band.
In the first case, see Fig.~\ref{DrudeAbs}, the reduced $C_{3v}$
symmetry manifests itself in the scattering matrix element $U_{1
{\bm k}', 1 {\bm k}}$, and, instead of Eq.~(\ref{shiftkk'}), one
should replace the elementary shift in the product $\left| M_{{\bm
k}' {\bm k}} \right|^2 {\bm R}_{{\bm k}' {\bm k}}$ by the
scattering induced shift
\[
{\bm R}^{(U)}_{{\bm k}' {\bm k}} = -  {{\rm Im}[ U^*_{1 {\bm k}',
1 {\bm k}} (\nabla_{\bm k} + \nabla_{{\bm k}'}) U_{1 {\bm k}', 1
{\bm k}}]\over \left| U_{1 {\bm k}', 1 {\bm k}} \right|^2}\:.
\]
Note that the main contribution to $U_{1 {\bm k}', 1 {\bm k}}$
dependent only on the difference $({\bm k}' - {\bm k})$ does not
result in the shift because the operator $(\nabla_{\bm k} +
\nabla_{{\bm k}'})$ nullifies such a term. Similar contribution to
the LPGE current has been calculated for Drude absorption in the
heavy-hole subband of a bulk $p$-GaAs,\cite{Beregulin88} where the
shift appeared due to interference of polar and deformation
scattering mechanisms on optical phonons.

For optical transitions going via $\Gamma_6$ valence-band states,
the elementary shift can be related to asymmetry of the interband
optical matrix elements in which case the asymmetry of the
interband scattering matrix element is neglected. We consider this
mechanism in more detail. Taking into account the symmetry
considerations we can present the interband matrix elements of the
operator ${\bm e}_\parallel \cdot {\bm p}$ (${\bm p}$ is the
momentum operator) between the conduction-band state $\Gamma_1$
and the valence-band states $\Gamma_6$ as
\begin{eqnarray} \label{epcv}
&&{\bm e}_\parallel\cdot{\bm p}_{c \Gamma_1;v, \Gamma_{6 x}}({\bm
k}) = {\rm i}
{\cal P} e_x + {\cal Q} (e_x k_y + e_y k_x)\:, \\
&&{\bm e}_\parallel \cdot {\bm p}_{c \Gamma_1;v, \Gamma_{6
y}}({\bm k}) = {\rm i}{\cal P} e_y + {\cal Q} (e_x k_x - e_y
k_y)\:, \nonumber
\end{eqnarray}
where ${\cal P}$ and ${\cal Q}$ are real constants. Multi-band
derivation of these contributions for GaN-based 2D structures is
given in Appendix~\ref{MatrElts}. The matrix element ${\cal Q}$ is
nonzero due to the presence of the interface in the
heterostructure. This approach is similar to description of
interface inversion asymmetry induced spin splitting in the
envelope-function method where $\delta$-functional terms are
introduced into the interband Hamiltonian, see
Refs.~\onlinecite{RoesslerKainz,GolubIvchenkoSiGe}. As a result,
we expect a strong dependence of ${\cal Q}$ on the confinement
size because ${\cal Q}$ is proportional to a product of the
electron and hole envelope functions at the interface, see
Eq.~(\ref{P&Q}). We expand $M_{{\bm k}'{\bm k}}$ and ${\bm
R}_{{\bm k}' {\bm k}}$ in powers of small parameters $\hbar
\omega/E_g$, $\bar{E}/E_g \ll 1$. In the same nonvanishing order
we should take into account $k$-dependence of the matrix element
${\cal P}$. This could be done by replacing in Eq.~(\ref{epcv})
the constant ${\cal P}$ by the function
\[
{\cal P}({\bm k}) = {\cal P}_0 \left( 1 - \frac{E_{\bm k}}{2 E_g}
\right)\:.
\]
The interband scattering $c \Gamma_1 {\bm k} \leftrightarrow v
\Gamma_{6x} {\bm k}'$ and $c \Gamma_1 {\bm k} \leftrightarrow v
\Gamma_{6y} {\bm k}'$ is described by the constant matrix elements
$U_{cv_x}$ and $U_{cv_y}$ assumed to be independent of ${\bm k}$
and ${\bm k}'$, with $\langle U_{c v_x}^2 \rangle = \langle U_{c
v_y}^2 \rangle$.

In order to carry out a reasonable estimation of the coefficient
$\chi'$ in Eqs.~(\ref{lpgexy}) we have used a simple
nonrelativistic ${\bm k} \cdot {\bm p}$ model coupling the $c
\Gamma_1$ and $v \Gamma_6$ band states. In this model the valence
band consists of two subbands, one with an infinite in-plane
effective mass and the other with the hole in-plane effective mass
coinciding with that for a conduction-band electron. Omitting the
details of derivation we present the final result
\begin{equation} \label{j_shift_normal_inc}
\chi' =   {2 \pi e^3 \over m_0^2 \omega} {\langle U_{cv_x}^2
\rangle {\cal P}_0 {\cal Q} \over E_g^3} N_s F_1\:,
\end{equation}
where $F_1 \equiv F [1] = \int\limits_{0}^{\infty} [f(E) - f(E +
\hbar \omega)] dE \left/ \int\limits_{0}^{\infty} f(E) dE
\right.$. Experiment shows a noticeable magnitude of the LPGE
current under normal incidence, see Fig.~\ref{fig4}.

\section{Discussion}\label{sec:Discussion}

In this Section we discuss the role of the contributions to the
total current and analyze the frequency and temperature
dependences of LPGE. An analytical estimation for the shift and
ballistic photocurrents at Drude absorption is readily available
for the photon energies $\hbar \omega$ small as compared to the
electron typical energy $\bar{E}$ which is relevant to our
experiments performed at room temperature and applying THz
radiation. In this case Eqs.~(\ref{j_shift_result}) and
(\ref{j_ball_result}) reduce to
\begin{eqnarray} \label{finalshift}
j_{\rm LPGE}^{({\rm shift})} &\sim& 2 \xi e z_{21}
{\eta_\parallel I \over \Delta_{21}} \:, \nonumber \\
j_{\rm LPGE}^{({\rm ball})} &\sim&  \xi e z_{21} {\eta_\parallel I
\over \Delta_{21}} \left( \hbar \omega \over 2 \bar{E}\right)^2\:.
\nonumber
\end{eqnarray}
Since $\hbar \omega /  \bar{E} \ll 1$ the shift contribution
dominates the LPGE. This estimation is confirmed by the study of
the frequency dependence of LPGE. Indeed,  while the shift
photocurrent  follows the frequency dependence of the absorbance
which, for Drude processes, is given by Eq.~\eqref{abs}, the
ballistic contribution has another frequency dependence because in
this case $j_{\rm LPGE}^{({\rm ball})} \propto \eta_\parallel
\omega^2 $.  The observed good agreement of the current and
absorption frequency behavior (see Fig.~\ref{fig3}) unambiguously
proves the dominating role of the shift contribution to the LPGE.
We note, that such a spectral behavior of the LPGE has been
previously observed in bulk III-V
semiconductors.\cite{Beregulin89p63}

We note that even for the opposite limit of $\hbar \omega /
\bar{E} \gg 1$ the LPGE due to Drude absorption is still dominated
by the shift contribution. In this case estimations yield ${\bm
j}_{\rm LPGE}^{({\rm shift})} = -(3/2) {\bm j}_{\rm LPGE}^{({\rm
ball})}$.

Now we analyze the temperature dependence. As follows from
Eq.~\eqref{j_shift_result}, the temperature dependence of the LPGE
current is given by $j_{\rm LPGE}^{({\rm shift})} \propto \xi \,
\eta_\parallel(T)$. Taking into account temperature behavior of
the Drude absorption [see Eq.~\eqref{Drude_absorbance}] we get
$j_{\rm LPGE}^{({\rm shift})} \propto \xi/\tau_p(T)$. Here we
again use the relation $\hbar \omega \ll \bar{E} \sim k_{\rm B}T$.
Variation of the temperature results in the change of the dominant
scattering mechanisms. Typically in the low-dimensional structures
by raising temperature we get a transition from  the
temperature-independent scattering rate by impurities
$1/\tau_p^{imp}$ to the fast-increasing rate due to scattering by
phonons $1/\tau_p^{ph}(T)$. As a result we have for the LPGE
current
\[
j_{\rm LPGE}^{({\rm shift})} \propto {\xi_{imp} \over
\tau_p^{imp}} + {\xi_{ph} \over \tau_p^{ph}(T)}.
\]
This expression demonstrates that despite the inverse mobility
$1/\mu(T) \propto 1/\tau_p^{imp}+1/\tau_p^{ph}(T)$ is a monotonous
function of temperature (see inset in Fig.~\ref{fig3}) and the
asymmetry factors $\xi$ are temperature-independent, the LPGE
current can have a minimum or even change the sign at temperature
increase. This is due to the interplay between two contributions
to the LPGE caused by different scattering mechanisms
characterized by substantially different values of $\xi_{imp}$ and
$\xi_{ph}$ possibly being even opposite in signs. Due to the fact
that only one time  is temperature dependent [$\tau_p^{ph}(T)$]
variation of temperature may result in the dominance of one or the
other contribution to the LPGE current.

\section{Conclusion}
In conclusion, we have studied the linear photogalvanic effect in
wurtzite GaN(0001)-based heterojunctions under excitation by
linearly and elliptically polarized light in the terahertz
frequency range. Experimentally, the properties of LPGE have been
revealed as a function of the incidence angle  and the
polarization state of radiation. A detailed point-group symmetry
analysis of the polarization properties shows an agreement between
the experiment data and the phenomenological theory of LPGE.
Phenomenologically, the effect  is described by two linearly
independent coefficients $\chi$ and $\chi'$, the former being
allowed by the C$_{3v}$, C$_{6v}$ and even C$_{\infty v}$
symmetries and the latter being nonzero only for the C$_{3v}$
symmetry. An observation of the photocurrent under normal
incidence determined by $\chi'$ unambiguously demonstrates a
reduction of the heterojunction symmetry from hexagonal C$_{6v}$
to trigonal C$_{3v}$. However, beginning from very small oblique
angles  first contribution to the photocurrent prevails over the
second one. The frequency dependence of the measured LPGE
photocurrent exhibits a Drude-like behavior. A microscopic theory
of LPGE under the intrasubband (Drude) absorption in the lowest
$e1$ electron subband has been developed. While calculating the
$\chi$-related photocurrent we have included virtual indirect
optical transitions via the higher subband $e2$, estimated both
the shift and ballistic contributions and showed that the first
one exceeds the other by a large factor $\propto (\bar{E}/\hbar
\omega)^2$. The dominating shift photocurrent repeats a Lorentzian
frequency dependence of the Drude-absorption coefficient, in
agreement with the experiment. The temperature dependence of the
photocurrent is governed by a product of the absorption
coefficient and the dimensionless parameter of asymmetry.
Microscopically, the contribution nonvanishing under normal
incidence arises taking into account indirect optical transitions
via virtual electronic states in the valence band.

\acknowledgments The financial support from the DFG and RFBR is
gratefully acknowledged. E.L.I. thanks DFG for the Merkator
professorship. Work of L.E.G. is also supported by ``Dynasty''
Foundation --- ICFPM and President grant for young scientists. The
high quality GaN samples were kindly provided by Hyun-Ick Cho and
Jung-Hee Lee from Kyungpook National University, Korea. We
gratefully acknowledge the support of the Stichting voor
Fundamenteel Onderzoek der Materie (FOM) in providing beam time on
FELIX.

\appendix

\section{Photocurrents in the rotated coordinate frame}

\begin{figure}[h]
\includegraphics[width=0.5\linewidth]{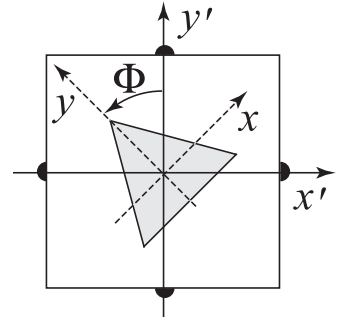}
\caption{Schematic representation of the orientation of the
crystallographic axes $x$, $y$ relative to the directions $x'$,
$y'$ along which the current is measured.} \label{fig:Phi-angle}
\end{figure}

In the Section IV.A we get phenomenological equations for LPGE at
normal incidence. We obtained that the photocurrent direction is
determined by the orientation of the electric vector with respect
to the crystallographic axes $x$ and $y$. Most convenient for this
effect is to investigate the current in the directions along and
perpendicular to one of the mirror reflection planes of the
$C_{3v}$ point group. However, in contrast to  III-V semiconductor
heterostructures where well determined in-plane crystallographic
orientation is naturally obtained by cleaving, in GaN/AlGaN
heterojunctions grown on the sapphire substrate this may be not
the case. Therefore we obtain here the LPGE for the arbitrary
orientation of contacts in respect to the crystallographic
directions. To describe this situation we introduce the angle
$\Phi$ between the in-plane directions $x'$ and $y'$ along which
the contacts are made and the crystallographic axes $x$ and $y$,
see Fig.~\ref{fig:Phi-angle}.
The symmetry considerations yield the expression for the
combination of $x'$ and $y'$ components of the photocurrent which
follows from  Eqs.~\eqref{lpgexy} for $E_z = 0$
\[
j_{x'} + {\rm i} j_{y'} = {\rm i} \chi' (E_{x'} - {\rm i}
E_{y'})(E_{x'}^* - {\rm i} E_{y'}^*) \exp({ 3 {\rm i} \Phi})
\]
or, equivalently,
\begin{eqnarray}
    \label{j_normal'}
&&j_{x'} =  \chi' (t_0 E_0)^2  \: \sin{(2 \alpha - 3\Phi)} \, , \\
&&j_{y'} =  \chi' (t_0 E_0)^2  \: \cos{(2 \alpha - 3\Phi)}  \, ,
\nonumber
\end{eqnarray}
where $\alpha$ is counted from $y'$: ${\bm E}(\alpha=0) \parallel
y'$.

Equations~(\ref{j_normal'}) demonstrate that the  azimuth angular
dependence for arbitrary orientation of contacts in respect to the
crystallographic directions remains the same but differs by the
phase shift $3\Phi$.

Finally we note that the photocurrent determined by the constant
$\chi$ in Eqs.~\eqref{lpgexy} is independent of the angle $\Phi$
because it is totally defined by position of the incidence plane
and polarization of light. Therefore Eqs.~\eqref{alpha-depend}
holds for any angle $\Phi$.

\section{Derivation of interband matrix elements}
\label{MatrElts}

Equation~(\ref{epcv}) can be obtained from the two-band electron
effective Hamiltonian
\begin{eqnarray} \label{Hcv}
&& {\cal H}_{c \Gamma_1;v, \Gamma_{6 x}}({\bm k}) =
\frac{\hbar}{m_0} ({\rm i} {\cal P} k_x + {\cal Q} k_x k_y)\:,
\\
&& {\cal H}_{c \Gamma_1;v, \Gamma_{6 y}}({\bm k}) =
\frac{\hbar}{m_0} \left[{\rm i} {\cal P} k_y + {\cal Q} \frac12
(k^2_x - k^2_y) \right]\:, \nonumber
\end{eqnarray}
by using the relation
\[
{\bm e} \cdot {\bm p}_{c \Gamma_1;v, \Gamma_6}({\bm k}) =
\frac{m_0}{\hbar} ({\bm e} \cdot \nabla_{\bm k}) {\cal H}_{c
\Gamma_1;v, \Gamma_6}({\bm k})\:.
\]

We describe the symmetry reduction from C$_{6v}$ to C$_{3v}$ by
introducing a $\delta$-functional perturbation $\hat{V}' \delta(z
- z_{\rm if})$, where $z_{\rm if}$ is the interface coordinate in
the $z$ axis and the operator $\hat{V}'$ transforms according to
the representation $B_1$ of the group C$_{6v}$ with the basic
function $y^3 - 3 x^2 y$ (here as before we assume that one of the
three mirror-reflection planes in the C$_{3v}$ group contains the
axes $y,z$ and is perpendicular to $x$). The perturbation
$\hat{V}'$ mixes the Bloch state $\Gamma_6$ with $\Gamma_5$ and
the Bloch state $\Gamma_3$ with $\Gamma_1$.

In order to find the quadratic-in-${\bm k}$ correction in
Eq.~(\ref{Hcv}) we calculate the third-order correction to the
Hamiltonian using the perturbation theory:\cite{PikusBir}
\[
\hat{\cal H}^{(3)}_{c;v} = \frac12 \sum_{s s'} \hat{\cal H}_{c;s}
\hat{\cal H}_{s,s'} \hat{\cal H}_{s'v} \Phi(ss')\:,
\]
\[
\Phi(ss') = \left[ \frac{1}{(E_c^0 - E_s^0)(E_c^0 - E_{s'}^0)} +
\frac{1}{(E_v^0 - E_s^0)(E_v^0 - E_{s'}^0)} \right]\:.
\]
Here the indices $c, v$ mean, respectively, the lowest conduction
and the highest valence bands of the symmetry $\Gamma_1$ and
$\Gamma_6$; $s, s'$ are other conduction or valence bands
different from $c$ and $v$ (the electron spectrum throughout the
Brillouin zone for bulk GaN can be found, e.g., in
Refs.~\onlinecite{bandstruct,matrixel}). $\hat{\cal H}_{n,n'}$ is
the matrix of the first-order matrix elements of the ${\bm k}
\cdot {\bm p}$ coupling or the interface mixing between the states
in the bands $n$ and $n'$; $E^0_n$ is the electron energy in the
band $n$ at the $\Gamma$ point. The dimension of the matrix
$\hat{\cal H}_{n,n'}$ is $N \times N'$ where $N, N'$ are
degeneracies of the bands $n$ and $n'$, respectively. Taking into
account the bands $s \Gamma_6$ ($s \neq v$), $s' \Gamma_5$ and
$s'' \Gamma_3$ we can present the coefficient ${\cal Q}$ in
Eq.~(\ref{Hcv}) for a structure with a single interface in the
following form
\begin{eqnarray} \label{PQ}
&& {\cal Q} = {\cal V} \delta(z - z_{\rm if}) \:,\\
{\cal V} &=& \frac{\hbar}{m_0}\ [P' (R' V_2 - V_1 R)\ \Phi(ss') -
V_3 P''R\ \Phi(s''s')]\ \:. \nonumber
\end{eqnarray}
The set of matrix elements is defined as follows:
\[
V_1 = \langle s, \Gamma_{6x}|\hat{V}' | s', \Gamma_{5x}
\rangle\:,\: V_2 = \langle v, \Gamma_{6x}|\hat{V}' | s',
\Gamma_{5x} \rangle\:,\]
\[\: V_3 = \langle c, \Gamma_1|\hat{V}'| s'',
\Gamma_3 \rangle\:,
\]
\[
P' = - {\rm i} \langle c, \Gamma_1| p_x | s,
\Gamma_{6x}\rangle\:,\:P'' = - {\rm i} \langle s'', \Gamma_3| p_x
| s', \Gamma_{5x}\rangle\:,
\]
\[
R = - {\rm i} \langle s', \Gamma_{5x}| p_x | v, \Gamma_{6x}
\rangle\:,\: R' = - {\rm i} \langle s', \Gamma_{5x}| p_x | s,
\Gamma_{6x} \rangle\:.
\]
For the band $s' \Gamma_5$ we choose the basis $| s, \Gamma_{5x}
\rangle$, $| s, \Gamma_{5y} \rangle$ which transforms under
operations of the C$_{3v}$ group as the coordinates $x$ and $y$,
i.e., as functions $| s', \Gamma_{6x} \rangle$, $| s', \Gamma_{6y}
\rangle$. While deriving Eq.~(\ref{PQ}) we took into account the
relation between matrix elements imposed by the symmetry, e.g.,
$\langle c, \Gamma_1| p_y | s, \Gamma_{6y}\rangle$ $= \langle c,
\Gamma_1| p_x | s, \Gamma_{6x}\rangle$, $\langle c, \Gamma_{5y}|
p_x | s, \Gamma_{6x} \rangle $ $= - \langle c, \Gamma_{5y}| p_y |
s, \Gamma_{6y}\rangle$ $= \langle c, \Gamma_{5x}| p_x | s,
\Gamma_{6x}\rangle$ etc.

It follows then that, for a heterojunction, the interband optical
matrix elements are indeed given by Eq.~(\ref{epcv}) where
\begin{equation}
\label{P&Q}
    {\cal P} =  P \int f_e(z) f_h(z) dz\:,
    \quad
    {\cal Q} = {\cal V} f_{e}(z_{\rm if}) f_{h}(z_{\rm if}) \:,
\end{equation}
$P = - {\rm i} \langle c, \Gamma_1 |p_x| v, \Gamma_{6 x} \rangle =
- {\rm i} \langle  c, \Gamma_1 |p_y| v, \Gamma_{6 y} \rangle$, and
$f_{e}(z)$, $f_{h}(z)$  are the electron and hole envelope
functions. 


\newpage

\begin{thebibliography}{80}

\bibitem{applphys} H. Linke (ed.), \emph{Ratchets and Brownian Motors: Basic Experiments and
Applications}, special issue Appl. Phys. A: Mater. Sci. Process. A
{\bf 75}, 167 (2002).


\bibitem{glass} A.\,M. Glass, D.\,von\,der Linde, and T.\,J.~Negran,
Appl. Phys. Lett. {\bf 25}, 233 (1974).

\bibitem{SturmanFridkin} B.\,I. Sturman and V.\,M. Fridkin, {\it The Photovoltaic and
Photorefractive Effects in Noncentrosymmetric Materials} (Gordon
and Breach,   Philadelphia, 1992).

\bibitem{10} E.\,L. Ivchenko, \textit{Optical Spectroscopy of
Semiconductor Nanostructures} (Alpha Science International,
Harrow, UK, 2005).

\bibitem{BelIvchStur} V.\,I. Belinicher, E.\,L.~Ivchenko, and B.\,I.~Sturman, Zh. Eksp. Teor.
Fiz. {\bf 83}, 649 (1982) [Sov. Phys. JETP {\bf 56}, 359 (1982)].

\bibitem{sinitsyn2}N.\,A. Sinitsyn, J. Phys.: Condens. Matter {\bf 20}, 023201 (2008).

\bibitem{APL05} W.~Weber, S.\,D.~Ganichev, Z.\,D.~Kvon, V.\,V.~Bel'kov, L.\,E.~Golub,
S.\,N.~Danilov, D.~Weiss, W.~Prettl, H.-I.~Cho, and J.-H.~Lee,
Appl. Phys. Lett. \textbf{87}, 262106 (2005).

\bibitem{ICPS06}  W.~Weber, S.\,D.~Ganichev, S.~Seidl, V.\,V.~Bel'kov, L.\,E.~Golub,
W.~Prettl, Z.\,D.~Kvon, Hyun-Ick~Cho, and Jung-Hee~Lee,
AIP Conf. Proc. \textbf{893}, 1311 (2007).

\bibitem{HeAPL2007} X.\,W.~He, B.~Shen, Y.\,Q.~Tang, N.~Tang, C.\,M.~Yin,
F.\,J.~Xu, Z.\,J.~Yang, G.\,Y.~Zhang, Y.\,H.~Chen, C.\,G.~Tang,
and Z.\,G.~Wang,
Appl. Phys. Lett. \textbf{91}, 071912 (2007).

\bibitem{TangApl2007} Y.\,Q.~Tang, B.~Shen, X.\,W.~He, K.~Han,
N.~Tang, W.\,H.~Chen, Z.\,J.~Yang, G.\,Y.~Zhang, Y.\,H.~Chen,
C.\,G.~Tang, Z.\,G.~Wang, K.\,S.~Cho, and Y.\,F.~Chen,
Appl. Phys. Lett. \textbf{91}, 071920 (2007).

\bibitem{ChoPRB2007} K.\,S.~Cho, C.-T.~Liang, Y.\,F.~Chen,
Y.\,Q.~Tang, and B.~Shen,
Phys. Rev. B \textbf{75}, 085327 (2007).

\bibitem{SSC07} W.~Weber, S.~Seidl,  V.\,V. Bel'kov, L.\,E.~Golub,
E.\,L.~Ivchenko, W.~Prettl, Z.\,D.~Kvon, Hyun-Ick~Cho,
Jung-Hee~Lee, and S.\,D.~Ganichev,
Solid State Comm. \textbf{145}, 56 (2008).

\bibitem{Nakamura} S. Nakamura and G.~Fasol, \textit{The Blue Laser Diode.
 GaN Based Light Emitters and Lasers},
(Springer, Berlin, 1997).

\bibitem{6} S.\,D.~Ganichev and W.~Prettl, \textit{Intense Terahertz Excitation of Semiconductors}
(Oxford University Press, Oxford, 2006).

\bibitem{Knippels99p1578}  G.\,M.\,H. Knippels, X.~Yan, A.\,M.~MacLeod,
W.\,A.~Gillespie, M.~Yasumoto, D.~Oepts, and A.\,F.\,G. van
der~Meer,
Phys. Rev. Lett. \textbf{ 83}, 1578 (1999).

\bibitem{Drude} K. Seeger, \textit{ Semiconductor Physics} (Springer, Wien, 1997).

\bibitem{Drude2}  N.\,V. Smith,
Phys. Rev. B \textbf{64}, 155106 (2001).

\bibitem{ChoApl2005} K.\,S. Cho, Tsai-Yu~Huang, Hong-Syuan~Wang, Ming-Gu~Lin, Tse-Ming~Chen,
C.-T.~Liang, Y.\,F. ~Chen, and Ikai~Lo,
Appl. Phys. Lett. \textbf{86}, 222102 (2005).

\bibitem{TangAPL2006} N.~Tang, B.~Shen, M.\,J.~Wang, K.~Han, Z.\,J.~Yang,
K.~Xu, G.\,Y.~Zhang, T.~Lin, B.~Zhu, W.\,Z.~Zhou, and J.\,H.~Chu,
Appl. Phys. Lett. \textbf{88}, 172112 (2006).

\bibitem{ThillosenAPL2006} N. Thillosen, Th.~Sch\"apers,
N.~Kaluza, H.~Hardtdegen, and V.\,A.~Guzenko,
Appl. Phys. Lett. \textbf{88}, 022111 (2006).

\bibitem{SchmultPRB2006} S. Schmult, M.\,J.~Manfra,
A.~Punnoose, A.\,M.~Sergent, K.\,W.~Baldwin, and R.\,J.~Molnar,
Phys. Rev. B \textbf{74}, 033302 (2006).


\bibitem{CingolaniPRB2000} R. Cingolani, A.~Botchkarev, H.~Tang, H.~Morkoc,
G.~Traetta, G.~Coli, M.~Lomascolo, A.~Di~Carlo, F.~Della~Sala, and
P. ~Lugli,
Phys. Rev. B \textbf{61}, 2711 (2000).

\bibitem{Litvinov}  V.\,I. Litvinov, Phys. Rev. B \textbf{68}, 155314 (2003).

\bibitem{Born_Wolf} M.~Born and  E.~Wolf, \textit{Principles of Optics} (Pergamon Press, Oxford, 1970).

\bibitem{Magarill_Entin}
L.\,I.~Magarill and M.\,V.~Entin, Fiz. Tverd. Tela
 \textbf{31}, 37 (1989) [Sov. Phys. Solid State \textbf{31}, 1299 (1989)].


\bibitem{book_Ivchenko_Pikus} E.\,L. Ivchenko and G.\,E.~Pikus,
    {\em Superlattices and Other Heterostructures. Symmetry and Optical Phenomena},
    Springer Series in Solid State Sciences, Vol.~110
    (Springer-Verlag, Heidelberg,  1995, 2nd ed. 1997).

\bibitem{sinitsyn} N.\,A. Sinitsyn, Q.~Niu, and A.\,H.~MacDonald, Phys. Rev. B {\bf 73},
075318 (2006).

\bibitem{ST_CPGE_orbital} S.\,A. Tarasenko, Pis'ma v ZhETF \textbf{85}, 216 (2007) [JETP Letters \textbf{85}, 182 (2007)].

\bibitem{ST_spin_orient}S.\,A. Tarasenko, Phys. Rev. B \textbf{73}, 115317 (2006).


\bibitem{Beregulin88} E.\,V.~Beregulin, S.\,D.~Ganichev, K.\,Yu.~Glukh, Yu.\,B.~Lyanda-Geller, and I.\,D.~Yaroshetskii, Fiz. Tverd. Tela
 \textbf{30}, 730 (1988) [Sov. Phys. Solid State \textbf{30}, 418 (1988)].


\bibitem{RoesslerKainz}U. R\"{o}ssler and J.~Kainz, Solid State Commun. \textbf{121}, 313 (2002).

\bibitem{GolubIvchenkoSiGe}L.\,E. Golub and E.\,L.~Ivchenko, Phys. Rev. B \textbf{69}, 115333 (2004).

\bibitem{Beregulin89p63} E.\,V.~Beregulin, S.\,D.~Ganichev, K.\,Yu.~Gloukh,
Yu.\,B.~Lyanda-Geller, and I.\,D.~Yaroshetskii,
Fiz. Tverd. Tela \textbf{31}, 115
(1989) [Sov. Phys. Solid State \textbf{  31},  63  (1989)].

\bibitem{PikusBir} G.\,L. Bir and G.\,E. Pikus, \textit{Symmetry and Strain-Induced Effects in Semiconductors} (Wiley, New York, 1974).


\bibitem{bandstruct} Zhongqin Yang and Zhizhong Xu, Phys. Rev.
B {\bf 54}, 17577 (1996).

\bibitem{matrixel} R. Beresford, J. Appl. Phys. {\bf 95}, 6216 (2004).


\end{thebibliography}
\end{document}